# On the theory of firm in nonlinear dynamic financial and economic systems

Dimitri O. Ledenyov and Viktor O. Ledenyov

*Abstract* – The new business paradigms originate a strong necessity to re-think the theory of the firm with the aim to get a better understanding on the organizational and functional principles of the firm, operating in the investment economies in the prosperous societies. In this connection, we make the innovative research to advance our scientific knowledge on the theory of firm in the conditions of the nonlinear dynamic financial and economic systems. We provide the definition of the firm and introduce the boundaries of the firm, defined by the barriers to entry, strategic boundaries, and limits to growth as in the theory of the firm. We use the econophysical evaluation and econometric estimation methods to analyze a full spectrum of the theories of the firm. We propose that the nonlinearities have to be taken to the consideration and the nonlinear differential equation have to be used to model the firm in the modern theories of the firm in the nonlinear dynamic financial and economic systems. We apply the econophysical approach with the dynamic regimes modeling on the bifurcation diagram as in the dynamic chaos theory with the purpose to accurately characterize the nonlinearities in the economic theory of the firm. We precisely characterize: 1) the dynamic properties of a combining number of the inventory orders in the stock as a function of the parameter $\lambda$ with the complex dependence on the forcing amplitude of the superposition of business cycles in the supply chain management theory; 2) the dynamic properties of the combining capital as a function of the parameter $\lambda$ with the complex dependence on the forcing amplitude of the total risk, which is defined by the superposition of the amplitudes of different risks, in the risk management theory; 3) the dynamic properties of the combining risky investments as a function of the parameter $\lambda$ with the complex dependence on the forcing amplitude of the standard deviation of return, which is defined by the superposition of amplitudes of oscillating asset classes in the modern portfolio theory. We introduce the *Ledenyov* firm stability theorem, based on the *Lyapunov* stability criteria, to precisely characterize the stability of the firm in the nonlinear dynamic financial and economic systems in the time of globalization.

PACS numbers: 89.65.Gh, 89.65.-s, 89.75.Fb

Keywords: theories of the firm, econophysics, econometrics, nonlinearities, nonlinear dynamic chaos theory, logistic function, logistic equation, bifurcation diagram, Ledenyov firm stability theorem, Lyapunov stability criteria, nonlinear dynamic financial and economic systems.

# Introduction

The ***theory of the firm*** is researched in the ***science of business administration***, which is a part of the ***science of economics*** in *Ueda (1904, 1937)*, *Mano (1968-1995)*. It is assumed that the one of the first attempts to create the *theory of the firm* was made by *Smith (1776)* during the analysis of the nature and causes of the wealth of nations. In the *science of business administration*, the *theory of the firm* seeks to explain the following three issues in *Spulber (2009)*: *1)* why firms exist, *2)* how firms are established, and *3)* what firm contributes to the economy. We can say that the theory *of the firm* mainly considers how the firms can use the *labor*, *capital* and *land* to operate effectively within the economies of the scale and scope. In fact, the *theory of the firm* includes a number of the theories on the *nature of the firm*, describing the origin, business behavior, organizational structure, and its relationship to the market in *Coase (1937)*, *Kantarelis (2007)*, and *Spulber (2009)*. Let us review the *theories of the firm*:

1. The ***neo-classical theory of the firm*** describes the various market structures, regulation issues, strategic pricing, barriers to entry, economies of scale and scope and even optimum portfolio selection of risky assets, and establishes the principle of profit maximisation, according to which profit is maximised, when marginal revenue is equal to marginal cost in the conditions of complete information. The theory does not allow for firm evolution in *Kantarelis (2007)*.

2. The ***transaction cost theory of the firm*** states that the people begin to organise their production in the firms, when the transaction cost of coordinating production through the market exchange in the conditions of the imperfect information, is greater than within the firm in *Coase (1937)*. It does not take into consideration agency costs or firm evolution, neither does it explain how vertical integration should take place in the face of investments in human assets, with unobservable value, that cannot be transferred in *Kantarelis (2007)*.

3. The ***managerial theory of the firm*** suggests that the managers would seek to maximise their own utility and consider the implications of this for firm behaviour in contrast to the profit-maximising case in *Baumol (1959, 1962)*, *Marris (1964)* and *Williamson (1966)*.

4. The ***principal–agent theory of the firm*** extends the ***neo-classical theory of the firm*** and ***managerial theory of the firm*** by adding agents to the firm, and it considers the friction due to asymmetric information between owners of firms and their stakeholders or managers and employees; the friction between agent and principal requires precise measurement of agent performance and the engineering of incentive mechanisms. The weaknesses of the theory are many: it is difficult to engineer incentive mechanisms, it relies on complicated incomplete contracts (borderline unenforceable), it ignores transaction costs (both external and internal), and it does not allow for firm evolution in *Spence and Zeckhauser (1971), Ross (1973), Kantarelis (2007)*.

5. The **behavioural theory of the firm** assumes that the groups of people participate in setting goals and making decisions on the production; inventory; market share; sales and profits in the firm, potentially creating conflicts. The theory proposes that the real firms aim to satisfy rather than maximize their results in agreement with the bounded rationality concept in *Cyert, March (1963)*.

6. The **evolutionary theory of the firm** states that the firm possesses unique resources: financial, physical, human and organizational. It sees the firm as a reactor to change and a creator of change for competitive advantage. The firm, as a creator of change, may cause creative destruction, which in turn may give birth to new industries and enable sectors of, or entire, economies to grow. The theory does not take to the account that the creative innovation process cannot be easily programmed within a firm or a nation in *Kantarelis (2007)*.

**Purpose and organizational structure of the firm**

Let us define the *purpose of the firm. Mano (1970)* defines the **purpose of the firm** as: "Therefore, the activity of the firm should be thought of as the process of producing the organizational utilities by combining the various cooperative activities and converting them into inducements and deriving the next contributions from the member mentioned above. In this case all the members want to sustain such cooperative activities to get more satisfaction (or inducements) and less sacrifice (or their cooperative activities). ***The actual purpose of leading principle of the firm can be abstractly said to be the maintenance and development of the firm itself. In other words, it can be said to be the maintenance of the balance of the organizational utilities and their increase.*** Unless the differential between the produced organizational utilities and the inducement derived from its utilities is equilibrium or positive, the firm will become bankrupt at some future time."

*Kantarelis (2007)* presents an interesting summary of his research findings on the **purpose of the firm**:

1. "The firm identifies a consumer need and develops/invents a recipe on how to satisfy that need;

2. The firm makes the right decisions with respect to making or buying inputs so that it delivers its recipe at the lowest possible cost;

3. The firm provides the best incentives to its stakeholders;

4. The firm constantly and deliberately evolves through the relentless pursuit of competitive, organization and strategic advantage."

Let us consider the organizational structure of *the firm* firstly, and then to describe the existing *theories of the firm* in the general framework of the *theory of the firm*. Barnard (1938, 1948, 1949, 1958) introduced the two types of organizations such as the *lateral organization* and *scalar (or hierarchical) organization*. According to *Mano (1970)*: "***The firm*** consists of not only stockholders, employees, and managers, but also creditors, government authorities, consumers and material suppliers:

1. *Stockholders* contribute a cooperative action to supply long-term capital in order to pursue the dividends, stock dividends, and the rise of stock value.

2. *Creditors* contribute a cooperative action to supply short-term capital mainly to earn interest.

3. *Consumers* contribute a cooperative action to supply cash in order to purchase goods or services.

4. *Government* authorities contribute a cooperative action to supply many conveniences in order to receive various taxes or donations, and material suppliers contribute a cooperative action to supply materials and facilities to get returns.

5. *Employees* and managers contribute the cooperative actions which combine the actions contributed by other members to produce organizational utilities in *Barnard (1938)* as large as possible and divide the utilities into various inducements or incentives as mentioned above in order to get contributions from members. They contribute such actions in order to get wages, utilizing right of employee benefit plans, social positions, honors and authority."

It makes sense to add that, during the period of evolution, *the firm* is defined by the ***boundaries of the firm***, which are mainly limited by such concepts as in *Chandler (2005)*:

1. *Barriers to entry*: the creation of business strategy and supporting management structure that enabled to create the intellectual property, technological innovation, supply and marketing and commercialization chains, establishing the barrier to entry in the certain market;

2. *Strategic boundaries*: the creation of business strategy to accurately define the scale and scope of the firm by focusing on the creation of the innovative products and services in the certain market;

3. *Limits to growth*: the creation of business strategy toward the careful strategic planning of the evolution of the firm in the certain market.

**Nonlinearities in theory of firm in nonlinear dynamic financial and economic systems**

Let us focus our attention on a full spectrum of various *theories of the firm*, which were created within the general framework of the *theory of the firm*, with the goal to define of *the identity of the firm* and try to understand the *functional and organizational principles of the firm*. *Lee (1975)* reviewed the *theories of the firm* and created the bird's eye view of relations between the different theories within the general framework of the *theory of the firm* as shown in Fig. 1.

*Lee (1975)* considered a big number of the different existing theories to model *the firm*:

1. **Economic theory** (the neoclassical microeconomics theory, including the mathematical programming theory, risk theory, uncertainty theory, dynamic theory, growth theory);

2. **Organization theory** (the classical organizational theory, neoclassical organizational theory, behavioural theory);

3. **Environmental theory**.

Let us focus our attention on the **economic theory of the firm**. *Lee (1975)* wrote: "With the aim of giving the microeconomics reality to be possibly used in practical business administration, there have been three orientation of progress (in the *economic theory of the firm*):

1. **Programming theory** with respect to fixed resources allocation;

2. **Uncertainty theory**, concerning with the imperfect knowledge;

3. **Dynamic theory**, related with the time."

Let us emphasis that the possible influence by the **nonlinearities** on the different economic and financial variables in the *programming theory*, which is positioned within the general framework of the *economic theory of the firm*, as it was suggested in *Lee (1975)*. For example, *Lee (1975)* made the following important comments: "***When the firm supplies an imperfect market with its product, the objective function is no more linear, because its marginal revenue is diminishing. This consideration for an imperfect market made the model for nonlinear programming required. Another requirement of nonlinear programming is found, when the problem of uncertainty is incorporated.***" Discussing the *uncertainty theory*, which is also positioned within the general framework of the *economic theory of the firm*, *Lee*

*(1975)* stated: "Imperfect knowledge is related with decision making in two kinds of respects: strategy generating and anticipating. … Approaches to decision making under uncertainty may be divided into two classes: probabilistic and game theoretic." Thus, let us summarize all the important research findings in the *economic theory of the firm* in *Lee (1975)*:

1. *Programming theory*: the nonlinear and linear programming approaches were discussed by *Lee (1975)*, and an appearance of the *nonlinearities* was connected with the problem of resources allocation in the condition of imperfect market;

2. *Risk theory:* the general understanding was that the *nonlinearities* may play an important role in the *risk theory*, but the origin of the *nonlinearities* was not well understood on that time and the use of the differential equation was not questioned;

3. *Uncertainty theory:* the general opinion was that the general understanding was that the *nonlinearities* may play an important role in the *uncertainty theory*, but the origin of *nonlinearities* was not well understood on that time and the use of the differential equation was not questioned;

4. *Dynamic theories*: the general position was that the *nonlinearities* may play an important role in the *dynamic theory*, but the origin of the *nonlinearities* was not well understood on that time and the use of the differential equation was not questioned.

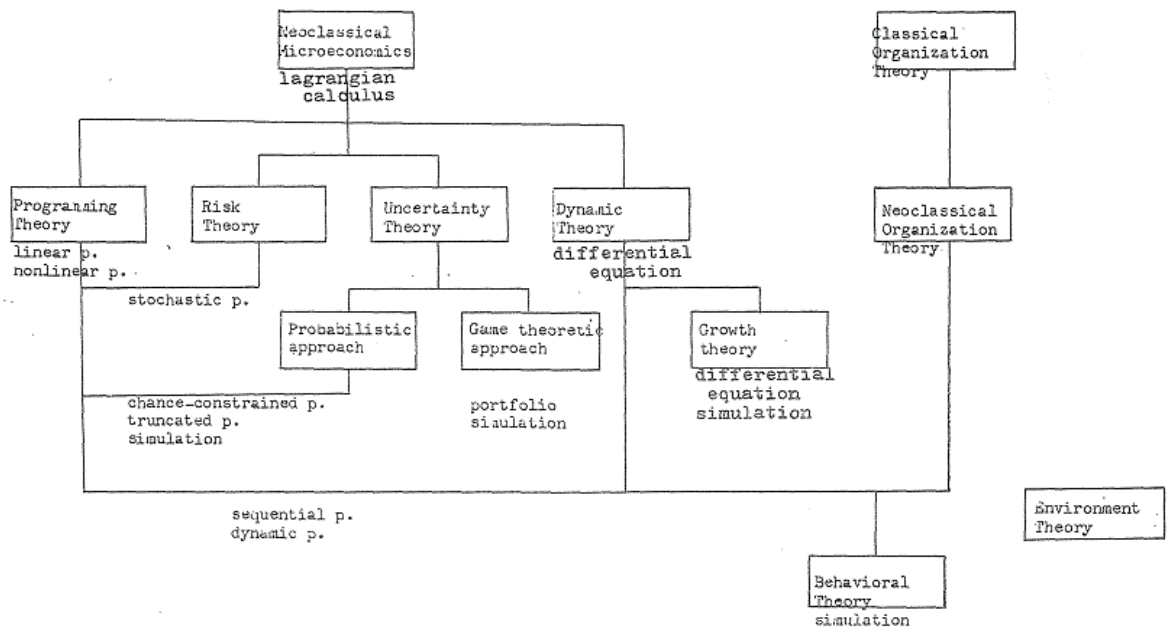

Fig 2. Bird's-eye view of theories of the firm.

* This is not a chronicle chart but only shows relation between theories.
* What is written under squares is a technique to solve the model of the theory in the square, which have been usually developed under domain of Management Science or operation research.

**Fig. 1.** *Bird's eye view of relations between different theories within general framework of theory of firm (after (Lee (1975)).*

*We propose that the nonlinearities have to be taken to the consideration and the nonlinear differential equations have to be used to model the firm, using the economic theory, organizational theory, environmental theory of the firm under an umbrella of the general theory of the firm, in the conditions of the nonlinear dynamic financial and economic systems.* Applying the **learning analytics** in *DeGioia (2012)*, *Faust (2012)*, *Rubenstein (2012)* together with the **integrative creative imperative intelligent conceptual co-lateral adaptive thinking** about the firm in *Martin (1998-1999, 2005-2006, 2007, 2008), Hart (2011)*, we decided to comprehensively analyze the *theories of the firm*, using the **econophysical evaluation** and **econometric estimation** methods; to model the *nonlinearities* in the *theories of the firm* and calculate such an important parameter as the *total stability of the firm*. We will use the extensive knowledge base, which was created during our research on the *nonlinearities in microwave superconductivity* in *Ledenyov D O, Ledenyov V O (2012)*, to complete the present research on the nonlinearities in the theory of the firm. Let us begin with the description of our research ideas on the **nonlinearities in the economic theory of the firm**, considering **the programming theory** with particular interest in **the theory of decision making** in the process of the **supply chain management**. *Mano (1970)* explains: "**The theory of decision making** in the firm comes to be one of the most important problems. We divide the decision making in the firm about the actual purposes of the firm into two cases: one is **the ordinary short-term actual purposes** and the other is **the strategic long-term ones**. First, the ordinary purposes of the firm are basically analyzed as follows. The demands of each member of the firm are inferred by the employees or executives and are offered to the top managers from each department in the forms of budget requests, staff increases, new plans, etc., together with the occupational demands of employees or executives. Here, I will look at the consumer's demands as an example. They are detected in marketing department, and offered to the executive committee as the demand for more progressive advertisement activity or new products, etc., together with the employees demands of the marketing department. At the same time, the demands of stock-holders are proposed to the executive committee by the financial department, for instance, as the necessity of constant dividends, together with another plan of financing.

We can recognize that each demand from each department is offered and arranged mutually, and at last some present actual purposes are decided. In this case, there are many demands; for example, mass production of a kind of commodity for cost decrease in the productive department and production of many kinds of commodities in the marketing department. The process of handling such demands is a very important problem. The way it is solved is that the resources (organizational utilities) are shared so as to fill the lowest level (or necessary level) of each department's demand, and no one special demand is pursued exclusively. It is a useful method in solving the above problem to not try to attain many purposes simultaneously, but rather to approach one of them at a time. At the same time, the firm tries to decide the work of both executives and employees, and to limit the necessary variety of information and the decision making of a person in order to get high accuracy. Also, the firm tries to make a person not pursue multiple purposes but to pursue one purpose at a time; this is the principle of division of labor in the firm. Moreover, the firm has various experiences or knowledge in the form of habits or standard operating procedures in *Cyert and March (1963)* but forward more accurate decision making; this system makes more accurate decisions in the firm than individuals. *The process of decision making for ordinary short-term purposes of the firm is such a process, so we think the communication system, management system or many short-term plans have been studied from this view point*.

On the other hand, the strategic long-term purposes are decided by many factors, especially the usage of synergy effects in *Ansoff (1965),* which depend upon the characters of organizational slacks in every firm, and the original innovative discoveries. These decisions should be finally made by the top management who are in a position to receive all important data or informations, and the employee and the executive join the decision-making process by supplying the necessary data. Here, organizational slack means the unused productive services, which were previously noted by *Charles Babbage* in *Babbage (1832)*, unused human resources in *Penrose (1966)*, especially unused human abilities that have developed with the growth of the human beings, and the excessive distributions of money or employees to commodities or businesses out of date or un-proportional distribution of expenses to income in *Drucker (1964)*. *Every firm has its own slack, so it is necessary to make suitable business policies according to the condition of the individual firm. Thus the strategic long-term purposes of the firms are mainly decided by the analysis of the organizational slack of each firm*."

Thus, let us consider the *decision making* problem, which is one of the most important problem, which has to be solved by the senior executives in any firm. It is a well known fact that the amplification phenomenon appears as a result of presence of the negative and positive feedback loops, which can originate the time delayed acceleration and multiplication mechanisms in the supply and distribution chains in the firm in *Mosekilde (1996, 1996-1997)*.Therefore, it makes sense to use the *econophysics* approach to research the supply chain in the economics, applying the computer modeling techniques, developed to research the nuclear chain reaction stability problem in the nuclear physics. In the case of the *supply chain management* in the firm, we believe that the accurate forecast of *a number of expected inventory* *orders in the stock* can only be calculated by taking to the account the existing *nonlinearities* in the *nonlinear dynamic financial and economic systems* by considering the dynamic behavior of the *inventory orders in the stock* in the firm in the conditions of the *nonlinear dynamic financial and economic systems*. Therefore, we propose that the dynamic behavior of the *inventory orders in the stock* during the *supply chain management* has to be closely approximated by **a simple nonlinear system model** with the physical characteristics, which strongly depend on the initial conditions of *the nonlinear dynamic financial and economic systems*. As we know the simple nonlinear financial systems do not possess the simple dynamical properties, because the nonlinear dynamic systems are usually characterized by the self-interactions, self-organizations, spontaneous emergence of order, dissipative structures and nonlinear cooperative phenomena in *Uechi, Akutsu (2012)*; *Mosekilde (1996, 1996-1997)*; *Kuznetsov (2001, 1996-1997)*. We have already learned from our previous publication in *Ledenyov D O, Ledenyov V O (2013)* that the physical behavior of a *simple nonlinear system* can be modeled, applying the **logistic function** in agreement with the nonlinear dynamic chaos theory in *Mosekilde (1996), Shiryaev (1998), Medvedeva (2000), Kuznetsov (2001)*

$$f_\lambda = \lambda x (1 - x).$$

Therefore, we decided to present the computed solutions of *logistic equation* on the **3D bifurcation diagram**, where the *expected number of orders* is plotted at the axis *Y* and the *parameter λ*, which has a complex dependence on the forcing amplitude of the mixed oscillations, generated by the business cycles, is plotted at the axis *X*. In our computer modeling, we computed the *1000 (one thousand)* iterations to plot the values of *the logistic function $f_\lambda^n (x_0)$* at the vertical axis *Y* and the values of the *parameter λ* at the horizontal axis *X*. In Fig. 2, the *3D bifurcation diagram* for accurate characterization of dynamic properties of a *combining number of the inventory orders in the stock* as a function of the *parameter λ* with the complex dependence on the forcing amplitude of the mixed business cycles in the nonlinear dynamic financial and economic systems is shown. Let us explain that we take to the consideration the *amplitudes*, *frequencies* and *phases* of the four main business cycles such as the *3 – 7 year Kitchin inventory cycle, the 7 –11 year Juglar fixed investment cycle, the 15 – 25 year Kuznets infrastructural investment cycle in Kuznets (1973) and the 45 – 60 Kondratieff long wave cycle in Kondratieff (1935)*, which are usually present in the nonlinear dynamic financial and economic systems in *Taniguchi, Bando, Nakayama (2008), Ikeda, Aoyama, Fujiwara, Iyetomi, Ogimoto, Souma, Yoshikawa (2012)*. Let us emphasis that the transition of the a combining number of the orders in the stock from **the stable state** to **the chaotic state** is realized through **the period doubling bifurcations** by which the *1 : 2* mode locking solution is transformed into the *2 : 4, 4 : 8, 8 : 16* solutions. It can be seen that, at the *high enough forcing amplitudes* of *the mixed business cycles*, the nonlinearities start to appear and the model begins to bifurcate, exhibiting the alterations between the high maximum and the low maximum. Let us explain that **the period two bifurcation** occurs at the parameter $\lambda \approx 3$. **The period four bifurcation** appears at the parameter $\lambda \approx 3,45$. At the further increase of the forcing amplitudes of *the mixed business cycles*, **the period doubling cascade of bifurcations** originates, and the model of simple nonlinear system transits to the state of **nonlinear dynamic chaos** with no regular periods at the parameter $\lambda_\infty = 3,5699$. We researched the complex dynamics of *the logistic function* up to the parameter $\lambda \approx 4$, obtaining the research results, which are in a good agreement with the **Sharkovsky-Yorke theorem** in *Sharkovsky (1964, 1965, 1986)*; *Li, Yorke (1975)*. We would like to note that the sharp expansion of the **chaotic attractor** is registered at the big enough forcing amplitudes of *the mixed business cycles*, resulting in a *crisis*, this research result fully complies with the research findings in *Grebogi, Ott, Yorke (1982)*. We observe the appearance of the *Kaplan-Yorke chaos* in the researched system similar to the case in *Mosekilde (1996)*.

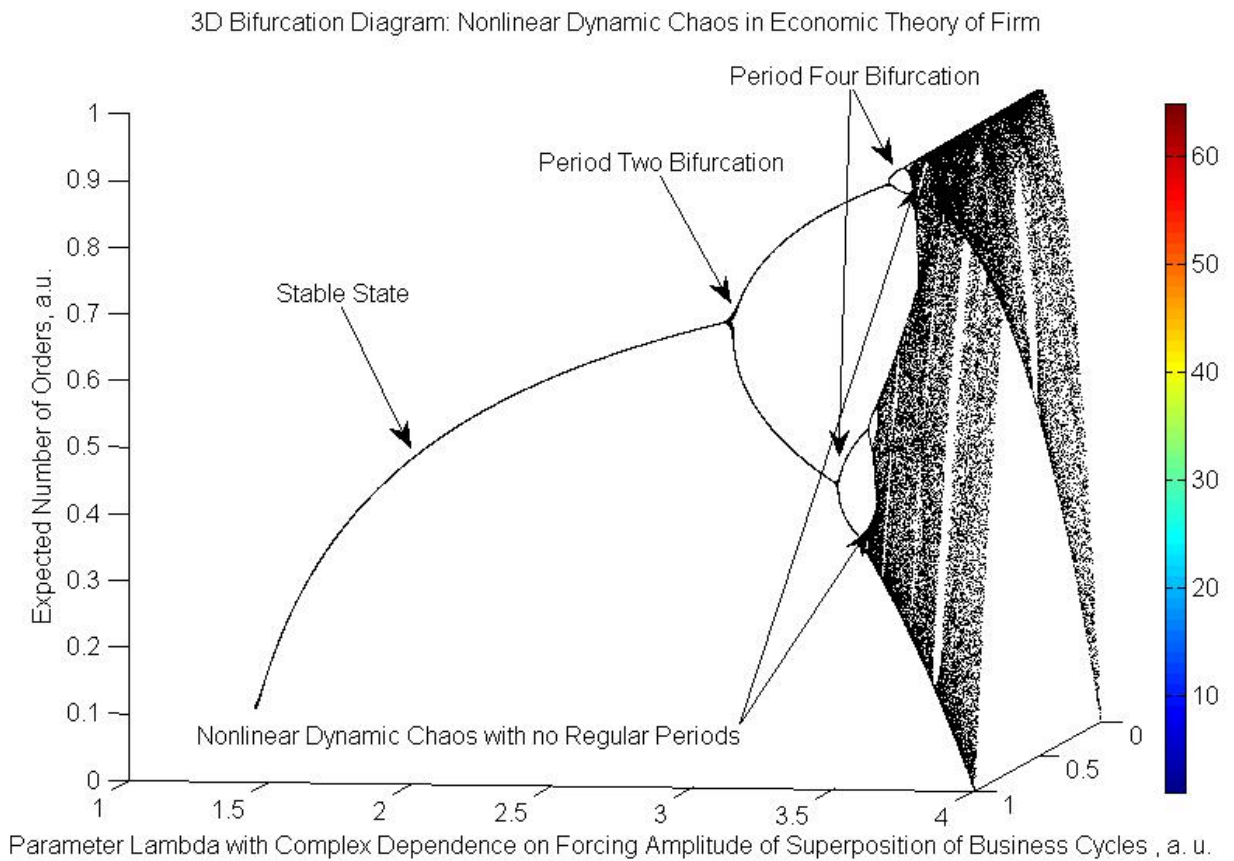

**Fig. 2.** *3D Bifurcation diagram for accurate characterization of dynamic properties of combining number of orders as a function of parameter λ with complex dependence on forcing amplitude of superposition of business cycles in nonlinear dynamic financial and economic systems (Ledenyov D O, Ledenyov V O (2013) Software in Matlab R2012).*

Going to the consideration our next research, we would like to report our new research results on the investigation of the influence by the **nonlinearities** appearance on the *capital allocation*, defined by the business decisions, which are based on the **risk management theory.**

Let us begin our explanation by describing the research results on the **integrated risk management framework** in *Meulbroek (2002)*: "*Integrated risk management* evaluates the firm's **total risk** exposure, instead of a partial evaluation of each risk in isolation, because it is the **total risk of the firm** which typically "matters" to the assessment of the firm's value and of its ability to fulfill its contractual obligations in the future. Furthermore, by aggregating risks, some individual risks within the firm will partially or completely offset each other (thereby reducing the total expense of hedging or otherwise managing those risks)." Thus, the *integrated risk management framework* can be used for formulating and designing a *risk management system* for the firm as explained in *Meulbroek (2002)*.

The three fundamental ways a company can implement its risk management objectives are in *Meulbroek (2002)*:

1) *Modifying the firm's operations*,
2) *Adjusting its capital structure*, and
3) *Employing targeted financial instruments*.

In Fig. 3, the **risk management system**, operating within the *integrated risk management framework*, is shown.

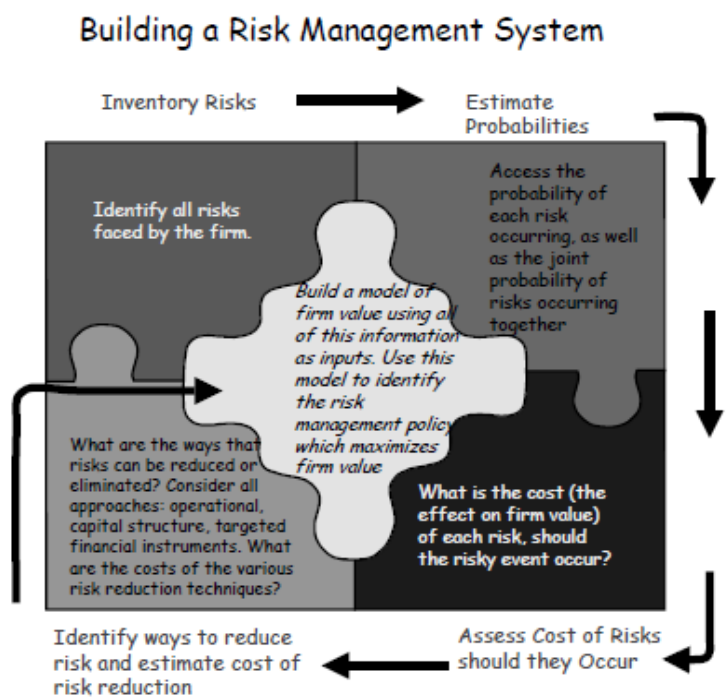

**Fig. 3.** *Risk management system* (*after Meulbroek (2002)*).

*Meulbroek (2002)* indentifies the **seven categories of risk**, including:

1. The *operational risk*,

2. The *product market risk*, *input risk*, *tax risk*, *regulatory risk*, *legal risk*, and *financial risk*. In accordance with *Meulbroek (2002)*, the firm's risk management team has to take to the account the probabilities of the *individual risks* occurring, to estimate the effect of a particular risk on firm value, then to calculate the *total risk* and to evaluate the effect of a *total risk* on the firm's valuation in *Meulbroek (2002)*.

In Fig. 4, the **total firm risk** and various **individual risks** are shown in *Meulbroek (2002)*.

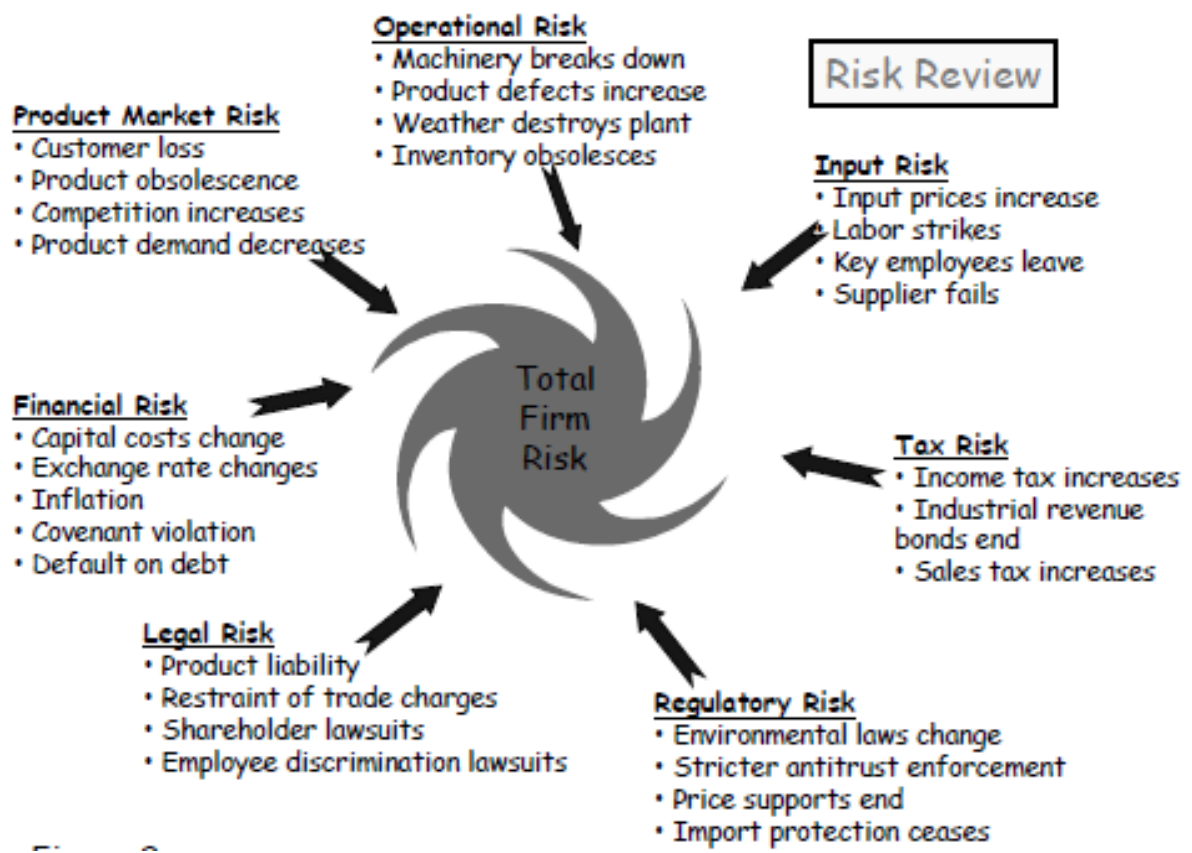

***Fig. 4.** Total firm risk after Meulbroek (2002)).*

*Meulbroek (2002)* concludes with a perspective on the future evolution of risk management in the firms and emphasizes the following facts: "**Corporate risk management** is evolving rapidly, but the practice of risk aggregation is not yet widespread. … So, as a practical matter, *integrated risk management* requires the unification (at least for the function of risk management) of previously separate institutional units. The firm, rather than the type of risk, provides a frame of reference."

In our view, the precise characterization of *the capital allocation by the firm* can only be made by taking to the account the existing *nonlinearities* in the *nonlinear dynamic financial and economic systems*. In analogy with the above considered case, we propose that, at the very first approximation, the dynamic behavior of the *combining capital in the firm* can be closely approximated by *a simple nonlinear system model*, which is characterized by the strong dependence of its initial conditions of *the nonlinear dynamic financial and economic systems*. We assume that every risk has the cyclic nature and can be characterized by periodic oscillations with the parameters such as the amplitude, frequency and phase, hence it can be described by the *wave function*. In our approximation, the *total risk* can be considered as a sum of all the risks faced by the firm during the capital allocation process at a given moment of time. Therefore, we can model a *simple nonlinear system*, applying the **logistic function** and compute the solutions of the *logistic equation*, representing the computed solutions on the **bifurcation diagram**, where *the expected capital* is plotted at the axis *Y* and the *parameter λ, which has a complex dependence on the forcing amplitude of the total risk* is plotted at the axis *X*. In other words, the solutions of *the logistic function $f_\lambda^n(x_0)$* are set at the vertical axis *Y* and the values of the *parameter λ* are set at the horizontal axis *X* after the computation of the *1000 (one thousand)* iterations.

$$f_\lambda = \lambda x (1 - x).$$

In Fig. 5, the 3D bifurcation diagram for the accurate characterization of the dynamic properties of the *combining capital* as a function of the *parameter λ* with the complex dependence on the *forcing amplitude of the total risk*, where the *total risk* is defined as a sum of the amplitudes of different periodic risks signals in the nonlinear dynamic financial and economic systems. We can certainly distinguish **the stable state** from **the chaotic state**, which is realized through **the period doubling bifurcations** by which the *1 : 2* mode locking solution is transformed into the *2 : 4, 4 : 8, 8 : 16* solutions. It must be noticed that, at the *high enough forcing amplitudes* of *the standard deviation of return*, the nonlinearities start to appear and the model begins to bifurcate, exhibiting the alterations between the high maximum and the low maximum. Let us repeat that **the period two bifurcation** occurs at the parameter *λ ≈ 3*. **The period four bifurcation** appears at the parameter *λ ≈ 3,45*. At the further increase of the forcing amplitudes of *the total risk*, the cascade of bifurcations originates, and the model of simple nonlinear system transits to the state of **chaos** with no regular periods at the parameter *λ∞ = 3,5699*. In this case, we also researched the complex dynamics of *the logistic function* up to the parameter *λ ≈ 4*, obtaining the research results, which are in a good agreement with the **Sharkovsky-Yorke theorem** in *Sharkovsky (1964, 1965, 1986)*; *Li, Yorke (1975)*. Once again, we would like to note that the sharp expansion of the ***chaotic attractor*** is registered at the big enough forcing amplitudes of *the standard deviation of return*, resulting in a *crisis*, this research result complies with the research findings in *Grebogi, Ott, Yorke (1982)*.

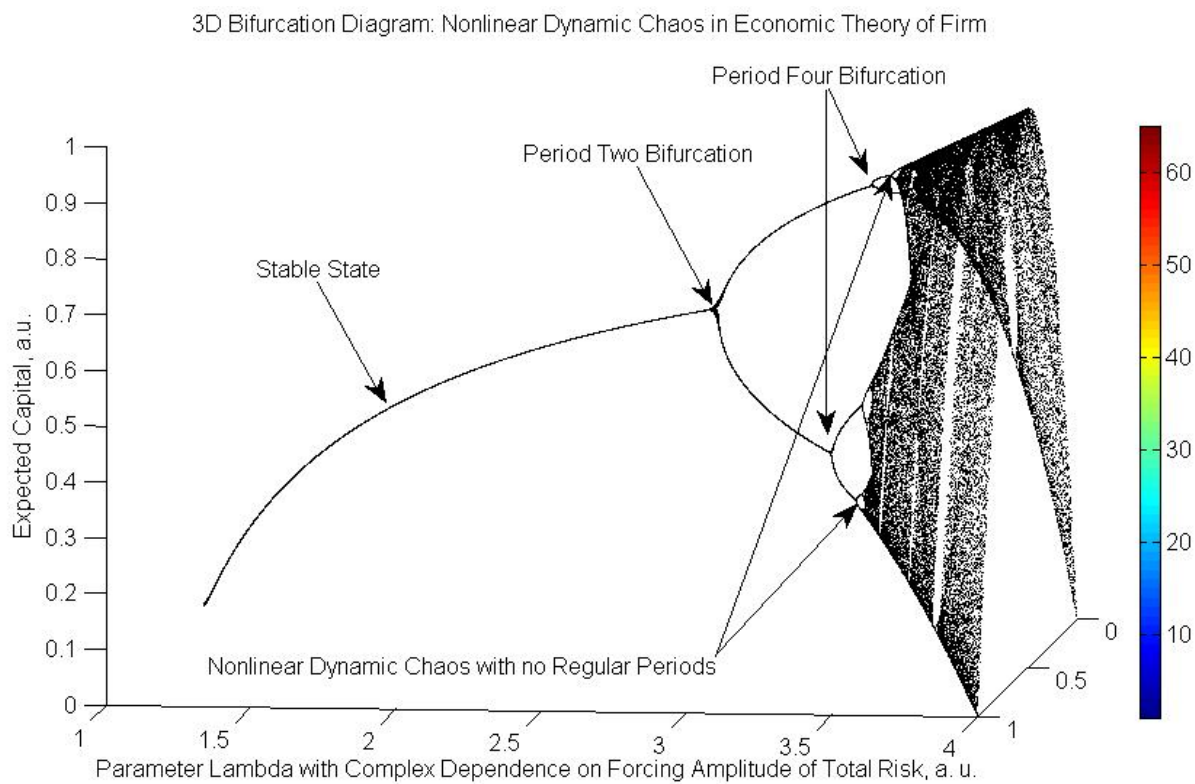

**Fig. 5.** *3D Bifurcation diagram for accurate characterization of dynamic properties of combining capital as a function of parameter λ with complex dependence on forcing amplitude of total risk , which is defined by the superposition of amplitudes of different risks, in nonlinear dynamic financial and economic systems (Ledenyov D O, Ledenyov V O (2013) Software in Matlab R2012).*

Presently, every firm is a money managing firm in *Kantarelis (2007), Hart (2011), Hart, Moore (1990), Lode Li (1986), Jensen, Meckling (1976)*. The management of risky investments is an important business task, which has to be solved by the senior executives in the firm. For example, the *Apple*, *Google*, *Facebook*, *Fujitsu, LG, Lenovo, Microsoft*, *Samsung* and some other corporations have accumulated the considerable financial resources (as a result of the increasing cash flow from the growing sales of their products and services as well as the increasing valuation of their shares at the stock exchanges in global capital markets), which have to be managed properly. Thus, in our opinion, the *nonlinearities* in the *nonlinear dynamic financial and economic system* must be considered during the process of accurate characterization of *the firm's investment portfolio*. In our view, this goal can be achieved, if the dynamic behavior of the *combining risky investments* in the *investment portfolio* will be closely approximated by *a simple nonlinear system model* with the physical characteristics, which strongly depend on the initial conditions of *the nonlinear dynamic financial system*. Using the *logistic function* in *Mosekilde (1996), Shiryaev (1998), Medvedeva (2000), Kuznetsov (2001)*, we can compute the solutions of *logistic equation*, presenting the outcomes on the *bifurcation diagram*, where *the expected return* is plotted at the axis $Y$ and the parameter $\lambda$, which has a complex dependence on the *forcing amplitude of the standard deviation of return* is plotted at the axis $X$. We can create the *bifurcation diagram* by plotting of the values of *the logistic function* $f_\lambda^n$ $(x_0)$ at the vertical axis $Y$ and the values of the parameter $\lambda$ at the horizontal axis $X$ after the computing of the *1000 (one thousand)* iterations

$$f_\lambda = \lambda x \left(1 - x\right).$$

In Fig. 6, the *3D bifurcation diagram* shows the transition of the *combining risky investments* in the *investment portfolio* from *the stable state* to *the chaotic state*, which is realized through *the period doubling bifurcations* by which the *1 : 2* mode locking solution is transformed into the *2 : 4, 4 : 8, 8 : 16* solutions. As it can be noticed, at the *high enough forcing amplitudes* of *the standard deviation of return*, the nonlinearities start to appear and the model begins to bifurcate, exhibiting the alterations between the high maximum and the low maximum. *The period two bifurcation* occurs at the parameter $\lambda \approx 3$. *The period four bifurcation* appears at the parameter $\lambda \approx 3,45$. The cascade of bifurcations originates, and the model of simple nonlinear system transits to the state of *chaos* with no regular periods at the parameter $\lambda_\infty = 3,5699$. In analogy to the above considered cases, we researched the complex dynamics of *the logistic function* up to the parameter $\lambda \approx 4$, obtaining the research results, which are in a good agreement with the *Sharkovsky-Yorke theorem* in *Sharkovsky (1964, 1965, 1986)*; *Li, Yorke (1975)*. The sharp expansion of the *chaotic attractor* occurs at the big enough forcing amplitudes of *the standard deviation of return*, resulting in a *crisis*, this research result complies with the research findings in *Grebogi, Ott, Yorke (1982)*. We would like to comment that, in our consideration, the choice of *the logistic equation* to describe a model of the *simple nonlinear system* was determined by the fact that this is a relatively well known mathematical approach to represent the complex physical behavior of a simple nonlinear system. It makes sense to note that we also continue our research work on the computer modeling with the application of other types of the nonlinear differential equations in *Sharkovsky, Maistrenko, Romanenko (1986), Strogatz (1994), Sauer et al (1996), Bunde, Havlin (2009)*. Moreover, we apply the advanced technique to quantify the chaos with the *Lyapunov exponents* in *Lyapunov (1950), Kuznetsov (2001), Cvitanovic, Artuso, Dahlquist, Mainieri, Tanner, Vattay, Whelan, Wirzba (2002)*.

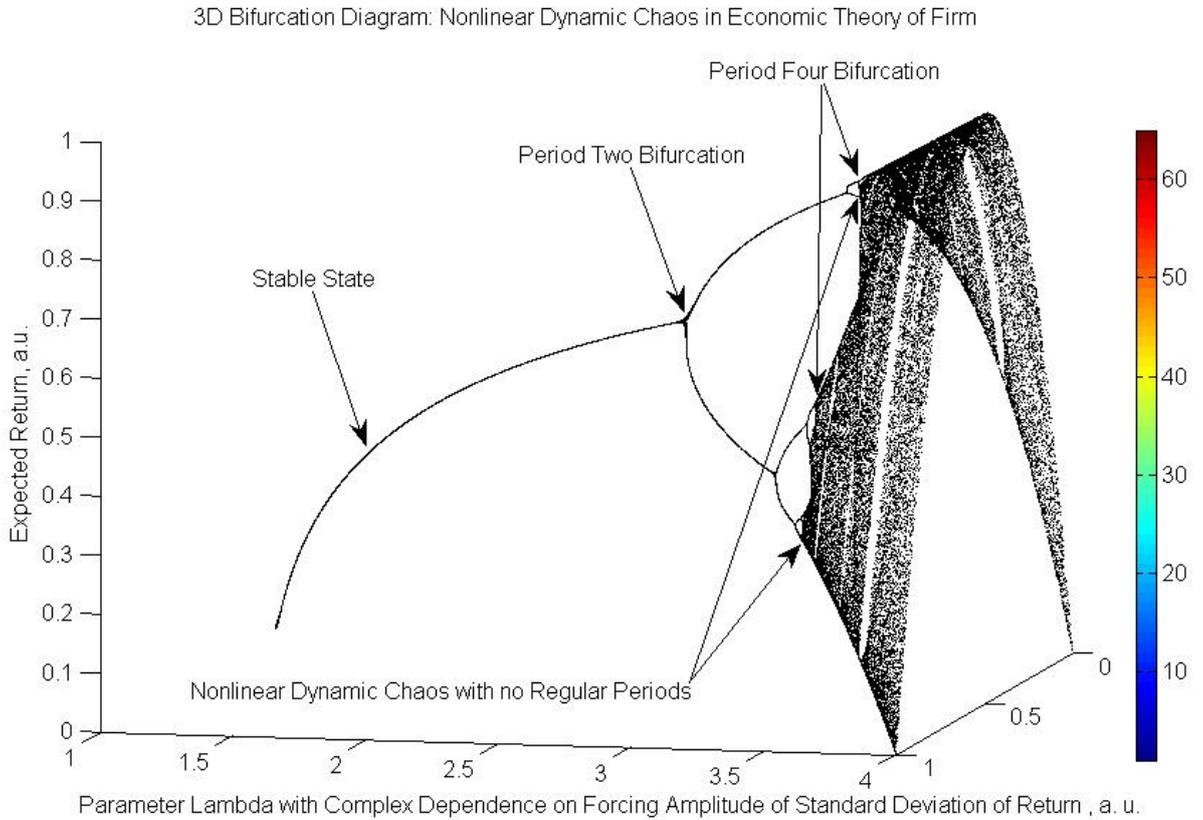

**Fig. 6.** *3D Bifurcation diagram for accurate characterization of dynamic properties of combining risky investments as a function of parameter λ with complex dependence on forcing amplitude of standard deviation of return, which is defined by the superposition of amplitudes of oscillating asset classes, in investment portfolio in nonlinear dynamic financial and economic systems (Ledenyov D O, Ledenyov V O (2013) Software in Matlab R2012).*

We propose the **<u>Ledenyov firm stability theorem</u>: The firm is stable in the case, when any pair of randomly selected variables in a set of the computed magnitudes of stabilities in the economic, organizational and environmental theories of the firm is stable, satisfying the Lyapunov stability criteria; namely the two randomly selected stabilities must have the two close trajectories at the start and continue to have the two close trajectories always.** We can see that the **Ledenyov firm stability theorem**, which is based on the **Lyapunov stability criteria** in *Lyapunov (1950), Mosekilde (1996), Kuznetsov (2001), Cvitanovic, Artuso, Dahlquist, Mainieri, Tanner, Vattay, Whelan, Wirzba (2002)*, can precisely characterize the *stability of the firm* and accurately forecast the business performance of the firm in the conditions of the *nonlinear* *dynamic financial and economic systems,* impacted by the globalization in *Stiglitz (2002)*, *Wolf (2005)*.

Let us provide some information on the *Lyapunov exponents* and explain that we used the *Lyapunov exponents*:

1) to determine the presence of the chaos in the system, and

2) to quantify the chaos.

*Cvitanovic, Artuso, Dahlquist, Mainieri, Tanner, Vattay, Whelan, Wirzba (2002)* explain: ""If all points in a neighborhood of a trajectory converge toward the same trajectory, the attractor is a fixed point or a limit cycle. However, if the attractor is strange, two trajectories that start out very close to each other separate exponentially with time, and in a finite time their separation attains the size of the accessible phase space."

$$x(t) = f^t(x_0) \ \ and \ \ x(t) + \delta x(t) = f^t(x_0 + \delta x(0))$$

The sensitivity of system to the initial conditions can be quantified as

$$\left| \delta x(t) \right| \approx e^{\lambda t} \left| \delta x(0) \right|$$

where $\lambda$ is called the **Lyapunov exponent**, representing the mean rate of separation of trajectories of system in *Lyapunov (1950), Cvitanovic, Artuso, Dahlquist, Mainieri, Tanner, Vattay, Whelan, Wirzba (2002)*. The *Lyapunov exponent* characterizes the average degree of increase of the distance between the two trajectories *Lyapunov (1950), Cvitanovic, Artuso, Dahlquist, Mainieri, Tanner, Vattay, Whelan, Wirzba (2002)*

$$\lambda = \lim_{t \to \infty} \frac{1}{t} \ln \frac{\left| \delta x(t) \right|}{\left| \delta x(0) \right|}.$$

The leading *Lyapunov exponent* can be computed as in *Lyapunov (1950), Cvitanovic, Artuso, Dahlquist, Mainieri, Tanner, Vattay, Whelan, Wirzba (2002)*

$$\lambda = \lim_{t \to \infty} \frac{1}{t} \ln \frac{\left| \mathbf{J}^t(x_0) \delta x(0) \right|}{\left| \delta x(0) \right|} = \lim_{t \to \infty} \frac{1}{2t} \ln \left| \hat{n}^T (\mathbf{J}^t)^T \mathbf{J}^t \hat{n} \right|,$$

where $\mathbf{J}_{ij}^t$ is the *Jacobian matrix*.

The time averaged *Lyapunov exponent* can be defined as in *Lyapunov (1950), Cvitanovic, Artuso, Dahlquist, Mainieri, Tanner, Vattay, Whelan, Wirzba (2002)*

$$\overline{\lambda_i(x_0)} = \lim_{t \to \infty} \frac{1}{t} \ln \left| \Lambda_i(x_0, t) \right|, \ \ i = 1, 2, ..., d.$$

We have to explain that we developed the software program to compute the **total stability of the firm** with the application of the unique algorithm, which calculates the *magnitude of stability* for every of the eight theories and sums up all the obtained magnitudes to get the *total stability of the firm*:

1. The **classical organization theory of the firm**;
2. The **neoclassical organization theory of the firm**;
3. The **transaction cost theory of the firm**;
4. The **managerial theory of the firm**;
5. The **principal–agent theory of the firm**;
6. The **behavioural theory of the firm**;
7. The **evolutionary theory of the firm.**
8. The **environment theory of the firm**.

In our research, during the calculation of the magnitudes of stabilities, we used the input-data fusion from the *Bloomberg terminal*. We precisely estimated the **total stability of the firm** for a number of the selected firms, which are traded at the *NASDAQ* and *NYSE*. We can complete the accurate characterization of the **total stability of the firm**, forecasting the firm (the company, corporation, financial institution, central bank) business performance in the short (*1* year) and long (up to *3* years) time periods. Presently, we continue our innovative research, focusing on the improvement of the computational algorithms.

## Conclusion

We live in the financial times in *Barber (2013)*, when the new business paradigms originate a strong necessity to re-think the existing theory of the firm in *Demsetz (1997), Ahuja (2007)* with the aim to get a better understanding on the organizational and functional principles of the firm, operating in the investment economies in the prosperous societies.

In this connection, we made the innovative research to advance our scientific knowledge on the theory of firm in the conditions of the nonlinear dynamic financial and economic systems. We provided the definition of the firm and explained the meaning of the boundaries of the firm, defining by the barriers to entry, strategic boundaries, and limits to growth as in the theory of the firm. We used the econophysical evaluation and econometric estimation methods to analyze a full spectrum of the theories of the firm.

We proposed that the nonlinearities have to be taken to the consideration and the nonlinear differential equation have to be used to model the firm in the modern theories of the firm in the nonlinear dynamic financial and economic systems. We applied the econophysical approach with the dynamic regimes modeling on the bifurcation diagram as in the dynamic chaos theory with the purpose to accurately characterize the nonlinearities in the economic theory of the firm.

We precisely characterized:

1) The dynamic properties of a combining number of the inventory orders in the stock as a function of the parameter $\lambda$ with the complex dependence on the forcing amplitude of the superposition of business cycles in the supply chain management theory;

2) The dynamic properties of the combining capital as a function of the parameter $\lambda$ with the complex dependence on the forcing amplitude of the total risk, which is defined by the superposition of the amplitudes of different risks, in the risk management theory;

3) The dynamic properties of the combining risky investments as a function of the parameter $\lambda$ with the complex dependence on the forcing amplitude of the standard deviation of return, which is defined by the superposition of amplitudes of oscillating asset classes in the modern portfolio theory.

We introduced the *Ledenyov firm stability theorem*, based on the Lyapunov *stability criteria*, to precisely characterize the *total stability of the firm* in the nonlinear dynamic financial and economic systems in the time of globalization in *Stiglitz (2002)*, *Wolf (2005)*.

## Acknowledgement

Authors are very grateful to the *Graduate School of Economics and Business Administration at Hokkaido University*, Sapporo, Hokkaido, Japan for presenting us with a wonderful opportunity to conduct the research with the highly innovative research papers, written by the *Japanese* scientists. We appreciate Prof. Geoffrey G. Jones from the *Harvard Business School Harvard University* in the *USA* for the thoughtful long discussion on the origin of chaos in the finances during our memorable meeting at the *Munk Centre for International Studies, Trinity College, University of Toronto* in Ontario, Canada in 2006. We also thank the *Harvard Business School, Harvard University* in Boston, *USA* for a presented opportunity to make our detailed analytical study on the current state of research on the theory of the firm, using an electronic collection of working papers at the *Harvard Business School*. The first author appreciates Prof. Janina E. Mazierska, *Electrical and Computer Engineering Department, School of Engineering and Physical Sciences, James Cook University*, Australia for an incredible opportunity to make the advanced innovative research on the modeling of nonlinear dynamic microwave resonant systems in the field of superconducting electronics during more than *12* years. The first author thanks a big group of the *Australian professors* from the *Victoria University* in Melbourne Australia for the exchanges by the scientific opinions on the economic and financial topics of mutual research interest in 2009. The second author would like to thank Prof. Erik Mosekilde from the *Center for the Modeling, Nonlinear Dynamics and Irreversible Thermodynamics* at the *Technical University of Denmark* in Lyngby in Denmark for the numerous thoughtful long discussions on the nonlinear dynamics, including the nature of chaos and hyper-chaos in the financial, economic and managerial systems in *1996-1997*. In addition, the second author would like to thank a group of the leading *Danish* professors for the stimulating philosophical discussions on the topics of the fundamental and applied economics as well as the encouraging discussions on the nonlinear dynamic chaos in the economics and finances, which had place at the *Copenhagen Business School*, *Copenhagen University*, *Henry George Library* in Copenhagen, Denmark and at the *Roskilde University* in Roskilde, Denmark in 1995-1997. The second author appreciates Prof. Harvey R. Campbell, *Fuqua Business School* for a most valuable opportunity to discuss the financial terminology in 2005, and thanks are also given to a group of the leading *American* professors from the *Fuqua Business School, Duke University* for the discussions on the application of different mathematical distributions in the finances during our numerous business meetings in Durham, North Carolina in the USA in 2005. The second author is deeply grateful to Profs. Roger L. Martin and John C. Hull from the *Rotman School of Management, University of Toronto* in Toronto, Canada for a wonderful opportunity to learn more about the integrative thinking, theory of the firm, and risk management in the finances and economics in North America at the *Rotman School of Management, University of Toronto* in Toronto, Canada in *1998-1999* and in *2005-2006*. The second author expresses his thanks Prof. Robert F. Engle III, *New York University* for the thoughtful scientific discussion on the modern portfolio, risk management and nonlinear dynamic chaos theories during our memorable meeting at the *Rotman School of Management, University of Toronto* in Toronto, Canada in 2006. The second author is grateful to Martin Wolf, *Chief Economic Commentator*, *Financial Times* for the insightful discussion on the globalization during our business meeting at the *Rotman School of Management, University of Toronto* in Toronto, Canada in 2006. Prof. Joseph Stiglitz, *Columbia University* is greatly acknowledged for the interesting debate on the topics of globalization and its discontents in 2002. The initial computing results on the nonlinear chaos dynamics in complex systems were obtained by the authors at the *Huskayne Business School*, *Calgary University* in Calgary, Alberta, Canada and at the *James Cook University* in Townsville, Australia in 1999-2002. The second author greatly benefited from and was influenced by the innumerable scientific advices on the fundamental and applied economics problems given by a number of the *Norwegian* professors during the scientific discussions at the *University of Oslo* in Oslo and our business meetings at the *Alexandra Hotel* in Loen, Norway in 1995-1997. The second author would like to appreciate Dr. Margaret Hankamp, *New York University*, who provided the enormous research support to the authors by setting up a computer center, searching for and obtaining some useful research book titles and discussing a number of the economic and financial issues in the course of our innovative research over the years. The second author would like to thank Hon. John Tefft, *US State Department* for his valuable strategic analysis on the recent developments in the *US* economy, finances and politics during our diplomatic meetings "without the ties," sharing our common passion for the *Jazz* music in Ukraine. Finally, Lionel Barber, *Editor-in-Chief, Financial Times* is appreciated for the discussions on the financial topics with more than one hundred global leaders, economists, financiers and academicians in the *Financial Times* in London in the *United Kingdom* in recent time.

*E-mail: dimitri.ledenyov@my.jcu.edu.au